
\input phyzzx \input epsf
%
%
%
%
%
\catcode`\@=11
\newfam\msbfam

\font\tenmsb=msbm10 \font\sevenmsb=msbm7 \font\fivemsb=msbm5
\textfont\msbfam=\tenmsb
\scriptfont\msbfam=\sevenmsb \scriptscriptfont\msbfam=\fivemsb
\def\Bbb{\relax\ifmmode\expandafter\Bbb@\else
 	\expandafter\nonmatherr@\expandafter\Bbb\fi}
\def\Bbb@#1{{\Bbb@@{#1}}}
\def\Bbb@@#1{\fam\msbfam\relax#1}
\catcode`\@=\active
%
%
%
%
%
\catcode`\@=11
\newfam\frakfam

\font\tenfrak=eufm10 \font\sevenfrak=eufm7 \font\fivefrak=eufm5
\textfont\frakfam=\tenfrak
\scriptfont\frakfam=\sevenfrak \scriptscriptfont\frakfam=\fivefrak
\def\frak{\n@expand\f@m\frakfam}
\catcode`\@=\active
%
%
%
%
%
\catcode`\@=11
\newfam\eusmfam

\font\teneusm=eusm10 \font\seveneusm=eusm7 \font\fiveeusm=eusm5
\textfont\eusmfam=\teneusm
\scriptfont\eusmfam=\seveneusm \scriptscriptfont\eusmfam=\fiveeusm
\def\script{\n@expand\f@m\eusmfam}
\catcode`\@=\active
%
%
%
%
%
\catcode`\@=11
\newfam\cyrfam

\font\tencyr=wncyr10
\font\sevencyr=wncyr7
\font\fivecyr=wncyr5
\def\cyr{\fam\cyrfam\tencyr}
\textfont\cyrfam=\tencyr \scriptfont\cyrfam=\sevencyr
	\scriptscriptfont\cyrfam=\fivecyr
\catcode`\@=\active
%
%
%
%
%

%
\font\eightcp=cmcsc8
\def\runningheads#1{\paperheadline={\iffrontpage\else
	{\hfil{\eightcp #1}\hfil}\fi}}
%
%
\catcode`\@=11
\def\sss{\scriptscriptstyle}

\def\A{{\frak A}}
\def\L{{\frak L}}
\def\M{{\frak M}}
\def\r{{\frak r}}
\def\s{{\frak s}}
\def\t{{\frak t}}
\def\u{{\frak u}}
\def\P{{\frak P}}
\def\Q{{\script Q}}

\def\R{\Bbb R}
\def\C{\Bbb C}
\def\H{\Bbb H}
\def\O{\Bbb O}
\def\K{\Bbb K}

\def\X{{\sss X}}
\def\Y{{\sss Y}}

\def\D{{\script D}}

\def\J{{\frak J}}
\def\Jx{\J^{\sss(\x)}}
\def\Jy{\J^{\sss(\y)}}
\def\Jz{\J^{\sss(\z)}}

\def\a{\alpha}
\def\b{\beta}
\def\x{\xi}
\def\p{\pi}
\def\px{\pi^{\sss(\x)}}
\def\pxs{\pi^{{\sss(\x)}*}\!}
\def\y{\eta}
\def\py{\pi^{\sss(\y)}}
\def\pys{\pi^{{\sss(\y)}*}\!}
\def\z{\zeta}

\def\pzs{\pi^{{\sss(\z)}*}\!}
\def\d{\delta}
\def\da{\delta_\alpha}
\def\db{\delta_\beta}

\def\w{\Omega}

\def\fullstop{\,\,.}
\def\komma{\,\,,}
\def\inn{\!\in\!}
\def\is{\!=\!}

\def\pro#1{\!\Buildrel
	\raise 4pt\hbox{\the\scriptscriptfont0 #1}\under\circ\!}
\def\Pro#1{\!\Buildrel#1\under\circ\!}

\def\nicefrac#1#2{\hbox{${#1\over #2}$}}
\def\frac#1/#2{\leavevmode\kern.1em\raise.5ex
		\hbox{\the\scriptfont0
         	#1}\kern-.1em/\kern-.15em
		\lower.25ex\hbox{\the\scriptfont0 #2}}

\def\wickcontract#1#2{
	\vtop{\baselineskip=0pt\lineskip=2pt
		\ialign{##&##\cr
			$#1$&$#2$\cr
			\hfill\vrule height 3pt depth 0pt
			\leaders\vrule height 1pt depth 0pt\hfill&
			\leaders\vrule height 1pt depth 0pt\hfill
			\vrule height 3pt depth 0pt\hfill\cr
		}}}

\def\arrover#1{
	\vtop{	\baselineskip=8pt\lineskip=2pt
		\ialign{  ##\cr
			$\longrightarrow$ \cr
                	\hfill ${\scriptstyle #1}$\hfill \cr   }}
	    }

\runningheads{M.~Cederwall and C.R.~Preitschopf, ``$S^7$ and
	$\widehat{S^7}$''}
\date={August, 1993}
\pubnum={\vbox{\hbox{G\"oteborg-ITP-93-34}
		\hbox{hep-th/9309030}}}
\titlepage
	\vskip 1cm
\title{\break{\fourteenpoint $S^7$ and $\widehat{S^7}$}}\vskip1cm
\author{Martin Cederwall\foot{e-mail tfemc@fy.chalmers.se}
{\rm and} Christian R.~Preitschopf\foot{e-mail tfecp@fy.chalmers.se}
\break\break}
\address{Institute for Theoretical Physics\break
	Chalmers University of Technology and University of G\"oteborg\break
	S-412 96 G\"oteborg, Sweden}
\vfill\abstract
We investigate the seven-sphere as a group-like manifold and
its extension to a Kac-Moody-like algebra. Covariance properties
and tensorial composition of spinors under $S^7$ are defined.
The relation to Malcev algebras is
established. The consequences for octonionic projective spaces
are examined.
Current algebras are formulated and their anomalies are derived, and
shown to be unique (even regarding numerical coefficients)
 up to redefinitions of the currents.
Nilpotency of the BRST operator is consistent with one particular expression
in the class of (field-dependent) anomalies. A Sugawara construction is
given.
\vfill\endpage
%
%
\REF\Cayley{A.~Cayley, \sl Phil.Mag. \bf 3 \rm (1845) 210.}
\REF\Zorn{M.~Zorn, \sl Abh.Math.Sem.Univ.Hamburg \bf 8 \rm (1930) 123.}
\REF\Schafer{R.D.~Schafer, ``An introduction to nonassociative algebras'',
	Academic Press,\nextline\indent New York (1966).}
\REF\Porteous{I.R.~Porteous, ``Topological geometry'', Cambridge University
	Press,\nextline\indent Cambridge (1981).}
\REF\Hurwitz{A.~Hurwitz, {\sl Nachr.Ges.Wiss.G\"ottingen} (1898) 309.}
\REF\BottMilnor{M.~Kervaire, \sl Proc.Nat.Acad.Sci.USA \bf 44 \rm (1958) 280;
	\nextline R.~Bott and J.~Milnor, {\sl Bull.Am.Math.Soc.}
	{\bf 64} (1958) 87.}
\REF\Adams{J.F.~Adams, \sl Ann.Math. \bf 75 \rm (1962) 603.}
\REF\Jacobson{N.~Jacobson, ``Lie algebras'', Wiley Interscience,
	New York (1962).}
\REF\Wolf{J.A.~Wolf, {\sl J.Diff.Geom.} {\bf 6} (1972) 317;
	{\bf 7} (1972) 19.}
\REF\Cartan{E.~Cartan and J.A.~Schouten, {\sl Proc.Kon.Wet.Amsterdam}
	{\bf 29} (1926) 803; 933.}
\REF\Husemoller{D.~Husemoller, ``Fibre bundles'', McGraw-Hill,
	New York (1966).}
%
%
\REF\GurseyTze{F.~G\"ursey and C.-H.~Tze, {\sl Phys.Lett} {\bf 127B}
	(1983) 191.}
\REF\Rooman{M.~Rooman, {\sl Nucl.Phys.} {\bf B236} (1984) 501.}
\REF\LukMinn{J.~Lukierski and P.~Minnaert,
	\sl Phys.Lett. \bf 129B \rm (1983) 392.}
\REF\softalg{F.~Englert, A.~Sevrin, W.~Troost, A.~Van Proyen
	and Ph.~Spindel,\nextline\indent{\sl J.Math.Phys.}
	{\bf 29} (1988) 281.}
\REF\Moufang{R.~Moufang, {\sl Abh.Math.Sem.Univ.Hamburg} {\bf 9} (1932) 207.}
\REF\JordanProj{P.~Jordan, \sl Abh.Math.Sem.Univ.Hamburg
	\bf 16 \rm (1949) 74.}
\REF\Freudenthal{H.~Freudenthal, \sl Advan.Math. \bf 1 \rm (1964) 145.}
\REF\Borel{A.~Borel, \sl Compt.Rend.Acad.Sci. \bf 230 \rm (1950) 1378.}
\REF\JordanJordan{P.~Jordan, \sl Nachr.Ges.Wiss.G\"ottingen \rm (1932) 569.}
\REF\JordanQuant{P.~Jordan, J.~v.~Neumann and E.~Wigner,
	\sl Ann.Math. \bf 35 \rm (1934) 29;\nextline
	M.~G\"unaydin, C.~Piron and H.~Ruegg,
	\sl Commun.Math.Phys. \bf 61 \rm (1978) 69.}
\REF\Albert{A.A.~Albert, \sl Ann.Math. \bf 35 \rm (1934) 65.}
\REF\Hopf{H.~Hopf, \sl Math.Ann. \bf 104 \rm (1931) 637,
	\sl Fund.Math. \bf XXV \rm (1935) 427.}
\REF\Sudbery{A.~Sudbery, \sl J.Phys. \bf A17 \rm (1984) 939.}
\REF\twistor{R.~Penrose and M.A.H.~McCallum
	\sl Phys.Rep. \bf 6 \rm (1972) 241
	\nextline\indent and references therein.}
\REF\htwistor{I.~Bengtsson and M.~Cederwall, \sl Nucl.Phys.
	\bf B302 \rm (1988) 81.}
\REF\tentwistor{N.~Berkovits, \sl Phys.Lett. \bf 247B \rm (1990) 45;\nextline
	M.~Cederwall \sl J.Math.Phys. \bf 33 \rm (1992) 388.}
\REF\jordanmech{M.~Cederwall, \sl Phys.Lett. \bf 210B \rm (1988) 169.}
\REF\supertwistor{A.~Ferber, \sl Nucl.Phys. \bf B132 \rm (1978) 55;\nextline
	T.~Shirafuji, \sl Progr.Theor.Phys. \bf 70 \rm (1983) 18.}
\REF\Malcev{A.I.~Malcev, {\cyr Mat.Sb.} {\bf 36(78)} (1955) 569;\nextline
	A.A.~Sagle, {\sl Trans.Am.Math.Soc.} {\bf 101} (1961) 426;\nextline
	H.C.~Myung, ``Malcev-admissible algebras'', Birkh\"auser,
	Basel (1986).}
\REF\KuzminII{E.N.~Kuzmin, {\cyr Algebra i Logika} {\bf 16} (1977) 424.}
\REF\OsipovCentral{E.P.~Osipov, {\sl Lett.Math.Phys.} {\bf 18} (1989) 35.}
\REF\OsipovSugawara{E.P.~Osipov, {\sl Phys.Lett.} {\bf 214B} (1988) 371.}
\REF\Defever{F.~Defever, W.~Troost and Z.~Hasiewisz,
	\sl J.Math.Phys. \bf 32 \rm (1991) 2285.}
%
%
\REF\Moody{R.V.~Moody, {\sl Bull.Am.Math.Soc.} {\bf 73} (1967) 217.}
\REF\GoddardOlive{P.~Goddard and D.~Olive, {\sl Int.J.Mod.Phys.} {\bf A1}
	(1986) 303.}
\REF\Schwinger{J.~Schwinger, {\sl Phys.Rev.Lett.} {\bf 3} (1959) 296.}
\REF\Sugawara{H.~Sugawara, \sl Phys.Rev. \bf 170 \rm (1968) 1659.}
\REF\Berkovitsstring{N. Berkovits, \sl Nucl.Phys. \bf B358 \rm (1991) 169.}
\REF\Eightconf{L.~Brink, M.~Cederwall and C.R.~Preitschopf,
	\sl Phys.Lett. \bf 311B \rm (1993) 76.}
\REF\hconf{M.~Cederwall and C.R.~Preitschopf, G\"oteborg-ITP-92-40,
	hep-th/9209107,\nextline\indent\sl Nucl.Phys. \bf B \rm in press.}
\REF\Holland{F.A. Bais, P. Bouwknegt, M. Surridge and K. Schoutens,
		\sl Nucl.Phys. \bf B304 \rm (1988) 348.}

%
%

\chapter{Preliminaries.}

This paper is devoted to an investigation of the seven-sphere as a manifold
equipped with group-like multiplication, and to its extension to a
Kac-Moody-like algebra. As is well known,
the seven-sphere is not a group manifold, but
shares a great number of properties with the group manifolds. It is
the parallelizability property that enables us to consider transformations
generated by vectors tangent to the seven-sphere. There are essentially two
routes to take when trying to generalize the Lie algebra concept. One,
which has been extensively explored in the mathematical literature,
is based on abandoning the Jacobi identities in favour of a weaker
structure, which leads to Malcev algebras. The other is to maintain
the Jacobi identities and give up the invariance of the structure
constants. It is this latter option that will be pursued in this paper,
for the simple reason that the multiplication rules defined by
Poisson brackets and commutators used in physics automatically
obey Jacobi identities.
We will also comment on the exact relation to Malcev algebras.

The paper is organized as follows. Section 1 gives a
brief summary of division algebras, specifically
aiming at octonions, which are an almost indispensable tool for
investigating the structure of the seven-sphere. Here we also
deal with the parallelizability properties of unit spheres related
to the division algebras. Section 2 uses the parallelizability to
define seven-sphere transformations and defines covariance properties
under these transformations. The current algebra is formulated and
its relation to Malcev algebras is established.
We discuss the implications for octonionic projective spaces, which are
naturally defined in our framework, and give a set of
homogeneous coordinates for $\O P^2$. Applications are examplified
by ten-dimensional twistor theory.
In Section 3 we consider
the Kac-Moody-like structure arising from the map $S^1\!\rightarrow\! S^7$.
We calculate the Schwinger terms, derive conditions for
quantum-mechanical nilpotency of the BRST operator and give a
Sugawara construction for an energy-momentum tensor. Sections 2 and 3
contain the results of this paper.
Section 4, finally, is devoted to a brief discussion of the results.

\section{Division algebras.}

The class of algebras of interest in this paper are the alternative
division algebras, especially the algebra $\O$ of octonions or
Cayley numbers [\Cayley,\Zorn]. As we will see, the properties of these
algebras are directly related to the corresponding algebras
of transformations as (properly defined) multiplication by an element
of unit norm.

An algebra $\A$ (not necessarily associative) is called a {\it division
algebra} (see \eg\ [\Schafer,\Porteous], which together with [\Zorn]
are the main sources
of this section) if left and right multiplication
$$L_ax\equiv ax\quad,\quad\quad R_ax\equiv xa	\eqn\leftright$$
have inverses (for $0\neq a\inn\A$). We will only consider division algebras
over the field $\Bbb R$ of real numbers. The existence of inverses
implies that there are no divisors of zero: $x\neq 0$ and $y\neq 0$
in $\A$ gives $xy\neq 0$.

An {\it alternative algebra} is an algebra
where the associator
$$[a,b,c]\equiv (ab)c-a(bc)	\eqn\associator$$
obeys the relation
$$[a,a,b]=0\fullstop			\eqn\alternative$$
This implies (consider $[a+b,a+b,c]$) that the associator alternates,
\ie\ changes sign under
any odd permutation of the entries. The alternativity implies a number
of useful relations, among which are the Moufang identities [\Zorn]
$$\eqalign{(axa)y&=a(x(ay))\komma\cr
	(ax)(ya)&=a(xy)a\fullstop\cr}\eqn\Mouf$$
The first of these equations is equivalent to
$$[a,xa,y]=-a[x,a,y]\fullstop\eqn\Mouff$$
One can prove that any alternative algebra $\A$
with a a unit element $1$ where any
nonzero element $x$ has an inverse $x^{-1}$ ($xx^{-1}\is 1\is x^{-1}x$)
is a division algebra. Namely, it follows from the Moufang
identity \Mouff\ that
$$[a^{-1},a,x]=0\eqn\Moufff$$
which means that left multiplication is invertible (\cf\ \leftright),
and analogously for right multiplication.

{\it Conjugation} of an element in $\A$ is defined as an anti-automorphism
$x\rightarrow x^*$: $(x^*)^*\is x$, $(xy)^*\is y^*x^*$, with
$$x+x^*\in\R\ni x^*x\fullstop\eqn\realconj$$
For convenience, we introduce the notation
$$[x]\equiv\half(x+x^*)\komma\quad\quad \{x\}\equiv\half(x-x^*)
	\komma\quad\quad |x|=(x^*x)^{\frac 1/2}
					\fullstop\eqn\realim$$
If $xx^*\neq 0$ we can obviously express the inverse in terms of the
conjugate element as $x^{-1}\is (xx^*)^{-1}x^*$. Thus, if there are
no zero divisors, an alternative $\A$ with conjugation
as above is a division algebra.

The class of finite-dimensional real division algebras is quite restricted,
they can be of dimensions 1, 2, 4 or 8 only [\Hurwitz,\BottMilnor,\Adams
]. If one also demands
that they be alternative, there are only four algebras left:
$\R$, the reals (dimension 1), $\C$, the complex numbers (dimension 2),
$\H$, the quaternions (dimension 4) and
$\O$, the octonions or Cayley numbers (dimension 8).
The algebra of octonions is unique
in that it is the only non-associative alternative division algebra.
We will refer to the above algebras as $\K_\nu$, where $\nu$ is the
dimension.

There is a number of equivalent ways to represent the multiplication
table of the octonions. The simplest one, to our opinion, is given
as follows. We chose an orthonormal basis
$$\O\ni x=\sum_{a=0}^7 x_ae_a=[x]+\sum_{i=1}^7x_ie_i\quad(e_0=1)
					\eqn\basis$$
and state the multiplication rule
$$e_ie_j=-\delta_{ij}+\sigma_{ijk}e_k\komma\eqn\multrule$$
where the structure constants $\sigma$ are completely antisymmetric
and equal to one for the combinations
$$(ijk)=(124),\,(235),\,(346),\,(457),\,(561),\,(672)\,\,{\rm and}\,\,(713)\,,
					\eqn\sigmatable$$
\ie\ we have
$$e_ie_{i+1}=e_{i+3}\komma\eqn\eeprod$$
where $e_{i+7}=e_i$. It is then easy to verify that the structure
constants for commutators and associators are given by
$$\eqalign{[e_i,e_j]&=2\sigma_{ijk}e_k\cr
	[e_i,e_j,e_k]&=2\rho_{ijkl}e_l\komma\quad\quad
	\rho_{ijkl}=-(^*\sigma)_{ijkl}=
		-{\nicefrac 1 6}\epsilon_{ijklmnp}\sigma_{mnp}
		\fullstop}\eqn\commass$$

\section{The seven-sphere. Parallelizability.}

The seven-sphere can be trivially represented as the set of unit octonions:
$$S^7=\{X\in\O\,|\,X^*X=1\}\fullstop\eqn\seveno$$
It is the unique compact and simply connected
non-group manifold
\foot{The non-compact spaces $SO(4,4)/SO(3,4)$
(topology $S^3\!\times\R^4$),
obtained from the split octonions [\Jacobson] and
$SO(8;\C)/SO(7;\C)$
(topology $S^7\!\times\R^7$)
also arise in the classification [\Wolf].}
to share with the group manifolds
the property of {\it global parallelizability} [\Cartan,\BottMilnor,\Wolf
,\Husemoller].
Of the spheres, only $S^{\nu-1}$, where $\nu$ is the dimension of
one of the alternative division algebras, are globally
parallelizable. The lower-dimensional spheres
are constructed analogously to above. There one has the isomorphisms
$S^1\approx U^1$ and $S^3\approx
SO(3)$, which leaves $S^7$ as the only non-group example.

Global parallelizability of a manifold $\M$ (in the following referred
to as just ``parallelizability'') means that there exist $m$ linearly
independent globally defined and nowhere vanishing vectorfields on $\M$,
where $m$ is the dimension of $\M$. Then the $m$ vectorfields can be
linearly combined to constitute an orthonormal basis of the tangent
space ${\script M}(X)$ at any point $X$ in $\M$. Letting the
orientation of this basis define parallel transport on $\M$, one
immediately obtains, since parallel transport is independent of path,
$$0=[\tilde\D_\mu,\tilde\D_\nu]=\tilde R_{\mu\nu}\komma\eqn\zerocurv$$
where $\tilde\D$ and $\tilde R$ are the covariant derivative
and the curvature tensor defined with respect to this parallel transport.
If we write
$$\tilde\D=\partial+\tilde\Gamma=\D-T\komma\eqn\covder$$
where $\D\is \partial+\Gamma$, $\Gamma$ being the metric connection,
we have the parallelizing connection $\tilde\Gamma\is \Gamma-T$.
$T$ is the {\it parallelizing torsion}. If one considers the covariant
derivative of the vielbein ${e_\mu}^a$ (roman letters $a,b,\ldots$
denote tangent space indices), one finds, since
$\D_\mu{e_\nu}^a\is 0$,
$$\tilde\D_\mu{e_\nu}^a=-{T_{\mu\nu}}^a\komma\eqn\torsiondef$$
or equivalently
$$\tilde\D_\mu{e_a}^\nu={T_{\mu a}}^\nu\komma\eqn\torsiondeff$$
which can be taken as the definition of torsion. 

\chapter{The seven-sphere as a transformational manifold.}

\section{From parallelizability to algebra.}

Suppose we have a manifold $\M$ of parallelizable type [\Wolf], \ie\
a direct product of group manifolds and seven-spheres (including
the complexified and non-compact versions). Then define
infinitesimal ``translations'' generated by
the tangent space covariant derivatives. The parallelizability
property assures that the translations form a closed algebra:
$$[\D_a,\D_b]=[{e_a}^\mu\D_\mu,{e_b}^\nu\D_\nu]=
	2{e_a}^\mu[\D_\mu,{e_b}^\nu]\D_\nu=
	2{e_a}^\mu {T_{\mu b}}^\nu\D_\nu=
	2{T_{ab}}^c\D_c	\komma\eqn\algebra$$
where we have utilized \zerocurv\ and the antisymmetry of the torsion
tensor [\Wolf].

The cases of group manifolds are trivial; there the parallelizing torsion
contains simply the structure constants, and is independent of
the location in $\M$. The Lie algebra of the group $\M$ is obtained.
The only ``non-trivial'' case is the
seven-sphere, and it is the only case where the torsion tensor
varies over the manifold. This statement contains exactly the same amount
of information as the statement that $\O$ is the only non-associative
alternative division algebra. We shall soon see the connection.

\section{Infinitesimal transformations. $S^7$ spinors.}

Let $\x\inn\O$ and let the seven-sphere be parametrized by the direction
of $\x$: $S^7\is\{{\x\over|\x|}\is X\inn\O\}$.
The tangent space at $X$ is spanned by the units
$\{Xe_i\}_{i\is 1}^7$. Considering the tangent space basis
at two infinitesimally separated points, we see that the parallel
transport of this basis is defined by an infinitesimal transformation
$$\da \x=\x\a\komma \quad\quad [\a]=0\fullstop\eqn\xtrans$$
The commutator of two such transformations can be calculated
explicitely:
$$\eqalign{[\da,\db]\x&\equiv\da(\db \x)-\db(\da \x)=(\x\a)\b-(\x\b)\a=\cr
	&=\x\bigl(X^*((X\a)\b)-X^*((X\b)\a)\bigr)=
	\d_{X^*((X\a)\b)-X^*((X\b)\a)}\x\fullstop\cr}\eqn\dadb$$
Here alternativity in the form \Moufff\ has been used.
The parameter $X^*((X\a)\b)-X^*((X\b)\a\is
2\{X^*((X\a)\b)\}$ of the transformation on the right
hand side is twice the parallelizing torsion [\GurseyTze,\Rooman].
In component notation,
$$T_{ijk}(X)=[(e^*_iX^*)(Xe_j)e_k]\komma\eqn\torsioncomp$$
which is completely antisymmetric in the three indices,
and \dadb\ can be written as
$$[\d_i,\d_j]=2T_{ijk}(X)\d_k\fullstop\eqn\didj$$
The variation $\d$ is indeed the parallelizing covariant derivative
of \algebra. It should be mentioned that the specific parallelizing
torsion used here is only one in a big family, parametrized by
the choice of left or right multiplication in \xtrans\ and by the
choice of the north pole [\Rooman]\foot{Here we have deduced the
parallelizing torsion in an indirect way, using \dadb\ and \algebra.
In reference [\LukMinn], it is shown how the torsionless and ``flat''
seven-spheres arise naturally as quotient spaces.}.
To our knowledge, the algebra
\didj\ was first considered in reference [\softalg].

Now the question arises how to transform other fields than $\x$.
One can not simultaneously interpret two fields as parametrizing the
seven-sphere. We want to introduce another boson
$\y$, ${\y\over |\y|}\is Y$ with some $S^7$
transformation rule, maintaining \didj.
This excludes the simplest candidate $\da Y\is Y\a$. The two fields are
bound to transform differently. The correct transformation rule turns
out to be
$$\da\y=X^*((X\y)\a)=(\y X^*)(X\a)=(\y(\a X^*))X\equiv\y\pro X\a
	\komma\eqn\spinortrans$$
where the equalities are derived from \Mouff.
By comparing with the tangent space basis introduced above,
one sees that the product \spinortrans\ indeed can be interpreted as
multiplication in the basis $(X,\{Xe_i\})$
at the point $X\inn S^7$. This multiplication
fulfills the same conditions as the ordinary multiplication at the
northpole $X\is 1$, and differs from it only by an associator.
We see that it is the non-associativity of $\O$ that is responsible
for the non-constancy of the torsion tensor (while the
non-commutativity accounts for its non-vanishing) and for the necessity
of utilizing inequivalent products associated with different points
$X\inn S^7$. We call this field-dependent multiplication
the $X$-{\it product}.

One should note that the transformation \spinortrans\ relies
on the transformation of the {\it parameter field} $X$ \xtrans,
while for group manifolds (and thus for the lower-dimensional
spheres $S^1$ and $S^3$ associated with $\C$ and $\H\,$) $\x$ and $\y$
transform independently. A consequence is that fermions cannot
transform without the presence of a parameter field, since
a fermionic octonion is not invertible.

We call a field (bosonic or fermionic) transforming according to
\spinortrans\ a {\it spinor under} $S^7$. Note that also the transformation
of $X$ can be written $\da X\is X\pro X\a$, and that the commutator
of variations is
$$[\da,\d]=\d\pro X\a-\a\pro X\d\komma\eqn\deltacomm$$
where $\d$ is thought of as an imaginary octonion: $\da\equiv[\a^*\d]$.

\section{$S^7$ tensor algebra.}

In order to examine covariance properties and tensorial
composition of spinors we will first examine the $X$-product
a little closer. We introduce the related commutators and associators:
$$\eqalign{[a,b]_\X&=a\pro X b-b\pro X a	\cr
	[a,b,c]_\X&=(a\pro X b)\pro X c-
	a\pro X(b\pro X c)	\fullstop\cr}\eqn\xcommass$$
Ordinary $*$ conjugation is still an anti-automorphism with respect to
the $X$-product. One also has
$$[a\pro X b]=[ab]\komma\quad\quad[a(b\pro X c)]=
		[(a\pro X b)c]\fullstop\eqn\xscalar$$
The Moufang identities \Mouf\ or \Mouff\ may be used to
express the $X$-associator in a number of ways.
The inverse $a^{-1}$ is also the inverse with respect to the $X$-product,
and alternativity, and thus also \Moufff\ holds.

Let $\r,\s,\ldots$ be $S^7$ spinors, \ie\ $\da\r\is\r\pro X\a$ \etc\
The generators $\d$ should be thought of as transforming in an
adjoint representation according to \deltacomm. Can this representation
be formed as a tensor product of spinor representations?
Due to the non-linearity, the answer is no. The current $\J$
(see the following section) is the
unique object to transform this way. The only reasonable candidate
for a spinor bilinear in the adjoint is $\{\r^*\!\pro X\s\}$ which does
not have good transformation properties\foot{See also section 3.2 on
BRST analysis. We want to emphasize that we do not yet have a full
representation theory.}.

On the other hand, consider a bilinear
$$K=\r\pro X\s^*\fullstop\eqn\Kdef$$
Had one used the ordinary multiplication ($X\is 1$), $K$ would not have
sensible transformation properties, but now also the product
itself transforms. We obtain
$$\da K=(\r\pro X\a)\pro X\s^*-[\r,\a,\s^*]_\X-\r\pro X(\a\pro X\s^*)=0
	\komma\eqn\Kscalar$$
where the middle term comes from transformation of the product,
which effectively {\it cancels the non-associativity}.
Let $K$ be scalar. The same cancellation of non-associativity
occurs in
$$\r'=K\pro X\r\quad :\quad\quad\da\r'=\r'\pro X\a\komma\eqn\rprimespinor$$
so that $\r'$ is a spinor. As a consequence, we can form spinors as
trilinears of spinors as
$$\u=(\r\pro X\s^*)\pro X\t\komma\eqn\trispinor$$
and in this way only.

\section{Current algebra.}

As mentioned in section 1, one of the main motivations for the
introduction of the field-dependent transformation rules is that the
Jacobi identities are fulfilled. Either this can be proven explicitely
by calculation, or it suffices to find a current that generates the
transformations \xtrans, \spinortrans\ by commutators or Poisson brackets.

Let $\px$ and $\py$ be the conjugate momenta of $\x$ and $\y$, \ie\
$$\{\x_a,\px_b\}=\d_{ab}=\{\y_a,\py_b\}\komma\eqn\conjmom$$
the curly brackets denoting Poisson brackets or commutators.
The transformation of $\x$ \xtrans\ is then easily derived from a generator
$\Jx=\{\pxs\x\}$ and that of $\y$ \spinortrans\ from
$\Jy=\{\pys\pro X\y\}$. $\Jx_\a\equiv[\a^*\Jx]$
fulfills the algebra \dadb\ with respect to the $\{\cdot\,,\cdot\}$
product, but $\Jy_\a$ does so only when $X$ transforms, \ie\
only in combination with $\Jx_\a$. The generator of the
simultaneous transformations \xtrans, \spinortrans\ is thus
$$\J=\{\pxs\x\}+\{\pys\pro X\y\}\fullstop\eqn\lmgen$$
In this expression, $\x$ is necessarily a boson, and $\y$ may be bosonic
or fermionic. Any number of fields can be introduced in $\J$ in the same
way as $\y$. Self-conjugated fermions ($\{S_a,S_b\}=\d_{ab}$) give
a contribution $\half S^*\!\pro X S$ to $\J$.

The transformation of any field $\phi$ is given by
$$\da\phi=\{\J_\a,\phi\}\fullstop\eqn\phitrans$$
Using \xcommass, the torsion tensor can be written as
$$T_{\a\b}(X)=\nicefrac 12[\a,\b]_\X\fullstop\eqn\simpletorsion$$
and the $S^7$ algebra \didj\ becomes
$$\{\J_\a,\J_\b\}=\J_{[\a,\b]_\X}\fullstop\eqn\gentrans$$
Note the exact analogy to $S^3\approx SO(3)$ obtained from $\H\,$,
where one has $\{\J_\a,\J_\b\}=\J_{[\a,\b]}$.

\section{Finite transformations.}

This section might be trivial, but the form of finite
transformations may not be so obvious when field dependence is involved.
A finite transformation is obtained by the limit procedure
$$\eqalign{&\phi\rightarrow\phi'=\lim_{{\sss N\rightarrow\infty}}
		\phi_{\sss N}\komma\cr
	&\phi_0=\phi\,,\quad\phi_{n+1}={\theta\over N}\{\J_\a,\phi_n\}+\phi_n
	\komma\cr}\eqn\finitelimit$$
where $\theta\inn\R$. A straightforward calculation shows that
$$\x\rightarrow\x'=\x\exp{(\theta\a)}=\x\w\komma\eqn\finxtrans$$
where $\w\inn S^7$. The corresponding calculation for a spinor $\r$
other than $\x$ is a little more involved, but a careful analysis shows
that all associators between $X$, $\a$ and $\r$ cancel, and the
finite transformation is
$$\r\rightarrow\r'=\r\pro X\w\komma\eqn\finspinortrans$$
and likewise for the current,
$$\J\rightarrow\J'=\w^*\!\pro X\J\pro X\w\fullstop\eqn\finjtrans$$
We will use this last equation later when looking at changes of
parameter fields in connection to projective spaces.

\section{Transformation of an arbitrary number of fields.}

The characterization of spinors under $S^7$ made above is not complete.
There are other ways for fields to transform than \spinortrans\ that
in a general treatment must be called spinorial.
Suppose that at least two bosonic fields transform under $\J$, and call
two of them $\x$ and $\y$ as before. The choice of $\x$ as parameter
field is arbitrary, one could as well have chosen $\y$, and there
is a way to move between the two forms of the current.
Namely, if we define
$$\tilde\J=((\J X^*)(XY^*))Y=(\J\pro XY^*)Y=(\J X^*)\pro YX
		\komma\eqn\jtildedef$$
we see that the roles of $\x$ and $\y$ have interchanged. We have
$$\tilde\J=\{\pys\y\}+\{\pxs\pro Y\x\}+\ldots\eqn\jtilde$$
Using only \jtildedef\ and not its explicit form \jtilde, one can show that
$\tilde\J$ fulfills the same algebra as $\J$, but with the torsion tensor
taken at the point $Y$ instead of $X$:
$$\{\tilde\J_\a,\tilde\J_\b\}=\tilde\J_{[\a,\b]_\Y}
		\fullstop\eqn\tildegentrans$$
Now, let there be yet another field $\z$ present, transforming the same
way as $\y$ under $\J$, and let $Z={\z\over |\z|}$.
The term $\Jz=\{\pzs\pro X\z\}$ in $\J$ does {\it not} change into
$\{\pzs\pro Y\z\}$ under the transformation \jtildedef.
Instead we obtain
$$\tilde\J=\{\pys\y\}+\{\pxs\pro Y\x\}+((\pzs(\z X^*))(XY^*))Y
		\fullstop\eqn\jtransformed$$
Note that the objects $ZX^*$ and $XY^*$ occurring in this formula
are $S^7$ scalars, according to \Kscalar, but that the remaining combination,
$YZ^*$ is not a scalar. We can visualize this in a linear diagram, where
$\x$ is connected with $\y$ and $\z$, but not $\y$ with $\z$ (figure 2, first
diagram). The transformation rule of $\z$ derived from \jtransformed\ is
$$\da\z=(\z X^*)((XY^*)(Y\a))\fullstop\eqn\twopath$$
Then the more complicated product in \twopath\ is thought of as a
product defined by going from $\z$ to the (new) parameter field $\y$
along the connections in this diagram. This principle is completely
generalizable to any connected ``tree diagram'' with arbitrary number of
points (fields) and arbitrary number of branches. Closed loops are forbidden;
due to non-associativity they lead to inconsistencies.

The path $\P[\r_n]$ from any field $\r_n$ to the parameter field $\r_0$ is
then uniquely determined, and letting it be the sequence
$$\P[\r_n]=\{\r_{n-1},\r_{n-2}\ldots,
		\r_1,\r_0\}\eqn\pathdef$$
(we enumerate the points in a tree diagram by the label $n$,
which is not an ordinary integer, but a set where subtraction by $1$,
\ie\ stepping towards the parameter point, is well-defined for $n\neq 0$,
but addition is not, due to possible branching),
we define the path-dependent product by
$$A\Pro{\P[\r_n]}B=(A\r_{n-1}^{-1})((\r_{n-1}\r_{n-2}^{-1})(\ldots
	((\r_2\r_1^{-1})((\r_1\r_0^{-1})(\r_0B)))\ldots))
			\komma\eqn\pathproduct$$
or by induction as
$$A\Pro{\P[\r_n]}B=(A\r_{n-1}^{-1})(\r_{n-1}\Pro{\P[\r_{n-1}]}B)
				\komma\eqn\pathprodII$$
which is easily seen to reduce to the product defined in \spinortrans\
for the case of a one-step and \twopath\ for a two-step path.
The generator of $S^7$ transformations is then
$$\J_\a=-\sum_k[\p_k^*(\r_k\Pro{\P[\r_k]}\a)]\komma\eqn\totalJ$$
and it fulfills \gentrans\ with $X\is\r_0/|\r_0|$.
Transformations are given as
$$\da\r_k=\r_k\Pro{\P[\r_k]}\a\fullstop\eqn\genertrans$$
One can also choose to view this as an ordinary multiplication by
another ($X$-dependent) unit octonion $\a_k$. In that case
one uses the properties of the path product to rewrite \totalJ\ as
$$\J_\a=-[\p_0^*\r_0\a]-\sum_{k\neq 0}[(\p_k^*\Pro{\r_{k-1}}\r_k)\a_k]
	\komma\eqn\alttotalJ$$
where $\a_k=1\Pro{\P[\r_k]}\a$, and obtains
$$\da\r_{k-1}=\r_{k-1}\a_k\fullstop\eqn\altgenertrans$$
Fermions, due to non-invertibility, can be assigned to endpoints
of the diagram only; no path may pass via a fermion.

Define the path from $\r$ to $\s$ as the composition
$$\P[\r,\s]=\P[\r]\P^{-1}[\s]\eqn\pathcomp$$
of the path of $\r$ followed by the reverse path of $\s$.
The invariance principle generalizing \Kdef, \Kscalar\ is
$$\da\,(\,\r\Pro{\P[\r\rightarrow\s]}\s^*\,)=0\komma\eqn\generscalar$$
which contains the irreducible amount of information that
$\da(\r\s^*)=0$ whenever $\r$ and $\s$ are connected by a link in the
diagram.

We conclude by the remark that change of parameter field along a link
in the diagram is a finite $S^7$ transformation, however not with
constant parameter. Consider $\J'=X\J X^*$, which has the property
of generating {\it left} multiplication on $X$. One can prove that
changing the parameter field from $X$ to the neighboring $Y$ amounts
to a finite $\J'$-transformation with (constant) parameter
$\w'=YX^*$. By using a modified form of \finjtrans, generated
by $\J'$, we obtain exactly \jtildedef.

\section{Octonionic projective spaces.}

This section will deal with the description of octonionic projective
spaces [\Porteous,\Moufang-\Albert] in terms of sets of homogeneous
coordinates modulo octonionic
transformations with $\O\!\setminus\!\{0\}\approx S^7\!\times\!\R^+$,
and to establish
relations with known coordinatizations.
There is a number of ``different''realizations of projective spaces:
from homogeneous coordinates [\Porteous],
from explicit sewing together coordinate patches [\Porteous,\Moufang],
from Jordan algebras [\JordanProj,\Freudenthal],
or as quotient spaces [\Borel].

An important feature is that $\K_\nu P^n$ can be described topologically as
the disjoint union of $\K_\nu^n$ and the space at infinity $\K_\nu P^{n-1}$
together with a map from the sphere at infinity $S^{n\nu-1}$ of $\K_\nu^n$
to $\K_\nu P^{n-1}$.
For the trivial case $\K_\nu P^1$, the maps are just constant maps from
$S^{\nu-1}$ to $\K_\nu P^0\is\{0\}$, that obviously can be taken as fibrations
with the ``group'' $S^{\nu-1}$ in the sense of this paper. One has
$\K_\nu P^1\is S^\nu$. For $\K_\nu P^2$, one has the maps from $S^{2\nu-1}$
to $\K_\nu P^1$, \ie\ the Hopf maps, or Hopf fibrations [\Hopf],
$$
\eqalign{
S^{1} \ & \arrover {S^0} \ S^1 = \R P^1 \cr
S^{3} \ & \arrover {S^1} \ S^2 = \C P^1 \cr
S^{7} \ & \arrover {S^3} \ S^4 = \H P^1 \cr
S^{15}\ & \arrover {S^7} \ S^8 = \O P^{1} \ . \cr
 }
\eqn\hopfmaps
$$
The map, together with scaling by positive real numbers, can be
used to obtain $\K_\nu P^{n-1}$ from its homogeneous coordinates in
$\K_\nu^n\!\setminus\!\{0\}$. This means that there is an equivalence
between the existence of $\K_\nu P^n$ and homogeneous coordinates for
$\K_\nu P^{n-1}$. This holds for $\nu\!\neq\!8$, all $n$,
and for $\nu\is8$, $n\!\leq\!2$.

We will show that the last of the Hopf fibrations \hopfmaps\ can be given a
formulation as a fibration with fiber $S^7$ in the sense of this paper,
\ie\ as identification of points on $S^{15}$ modulo infinitesimally
generated $S^7$ orbits without fixed points. In view of the relation to
homogeneous coordinates, the same holds for the map to $\O P^1$ from
its homogeneous coordinates.

We start with $\O P^1 (\approx\!S^8)$, whose standard atlas consists
of the the two charts
$$\eqalign{&(1,y_1)\cr
	&(x_2,1)\cr}\eqn\onecharts$$
with the overlap equation $x_2\is {y_1}^{-1}$ where both charts are
valid. The standard homogeneous coordinates of $\O P^1$ are a pair
of octonions $(\x,\y)$ defined modulo (right) multiplication with
the same octonion:
$$
(\x,\y)\approx(\x\w,\y\w)\fullstop
\eqn\onehom
$$
The consistency with \onecharts\ is seen by chosing $\w\is\x^{-1}$
or $\w\is\y^{-1}$. One has to be careful, however. The transformations
of \onehom\ do not close to an algebra (see section 2.2), so repeated
use of them does not give an equivalence class of points corresponding
to the same point in $\O P^1$. A basepoint has to be chosen (preferrably
in one of the forms of \onecharts). A more natural way,
at least from our point of view, would be
to define the homogeneous coordinates modulo $S^7$ transformations
(and a real scale). We then have the transformations according to figure 1:
$$
\eqalign{&(\x,\y)\approx(\x\w,\y\pro X\w)\komma\quad\x\neq 0\komma\cr
	&(\x,\y)\approx(\x\pro Y\w',\y\w')\komma\quad\y\neq 0\fullstop\cr}
\eqn\onehommod
$$
The price being paid
for the algebra structure for the variables parametrizing the $S^7$
fiber is that we cannot describe an equivalence class by
only one of the transformations \onehommod, since for the
$X$-($Y$-)product is undefined at $\x\is 0$($\y\is 0$).
Of course the transformations
are equivalent for $\x,\y\neq 0$: $\w= Y^* (Y\pro X\w')$, since the
associated currents are related according to \jtildedef.
The mapping from the homogeneous coordinates (modulo $\R^+$) to
$\O P^1$ is a topologically equivalent modification of the Hopf map
$S^{15}\!\rightarrow\!S^8$ [\Hopf],
that now has been turned into identification of points on {\it infinitesimally
generated} $S^7$ orbits (see also section (2.8) for a physical motivation).

The reason that the traditional homogeneous coordinates for $\O P^1$
exist, is that the specific $\w$'s taking $(\x,\y)$ to \onecharts\
satisfies $[\w,\x,\y]\is 0$. Trying the same procedure for $\O P^2$
is bound to fail~-- the atlas
$$\eqalign{&(1,y_1,z_1)\cr
	&(x_2,1,z_2)\cr
	&(x_3,y_3,1)\cr}\eqn\twocharts$$
with the overlap equations
$$\matrix{&x_2=y_1^{-1}\hfill&&z_2=z_1y_1^{-1}\hfill\cr
	&x_3=x_2z_2^{-1}\hfill&y_3=z_2^{-1}\hfill&\cr
	&&y_1=y_3x_3^{-1}\hfill&z_1=x_3^{-1}\hfill\cr}\eqn\overlap$$
(consistency is easily checked)
can not be reached from $(\x,\y,\z)$ by uniform right octonionic
multiplication, due to non-associa\-tiv\-ity. We need the $S^7$
transformations, fulfilling Jacobi identities and thus effectively
associative. Any set of coordinate patches
of the generic type \twocharts\ resulting from identifying points on
$S^7$ orbits in some coordinates {\it within one specific
diagram} is automatically consistent in the regions where the overlap
equations apply~-- this follows from the composition properties
of the $S^7$ transformations.

Let us now try to construct homogeneous coordinates. We choose
a linear
diagram of the variables $(\x,\y,\z)$, with $\x$ as parameter field
in the middle (figure 2, first diagram). Points on $S^7\times\R^+$ orbits are
identified as
$$(\x,\y,\z)\approx(\x\w,\y\pro X\w,\z\pro X\w)\eqn\homtwotry$$
(it is easily checked that \twocharts\ with \overlap\ holds).
This map has a problem for $\x\is 0$. In the twentythree-sphere
$|\x|^2+|\y|^2+|\z|^2=1$, approaching the fifteen-sphere
$\x\is 0$ from the seven-sphere direction $X$ gives the $\O P^1$ charts
$$\eqalign{&(0,1,\z\pro X\y^{-1})\cr
	&(0,\y\pro X\z^{-1},1)\cr}\eqn\twosubone$$
and the orbits are not well defined on $\x\is0$ unless we explicitely
specify the value of $X$ there. This choice has to be consistent with
the transformations, and we note that, whatever prescription is used,
$$\lim_{\x\rightarrow 0}\pi(\x,\y,\z)\neq\pi(0,\y,\z)\eqn\discont$$
for some directions, where $\pi$ is the map $S^{23}\rightarrow\O P^2$.
We arrive at a discontinuous fibration as a generalization of the
Hopf map. We make the choice of prescription in \homtwotry:
$$X=\cases{\x/|\x|\komma	&if $\x\neq 0$\ ;			\cr
	\y/|\y|\komma 		&if $\x=0$, $\y\neq 0$\ ;		\cr
	\z/|\z|\quad \hbox{(or any $X'\inn S^7$)}\komma
				&if $\x=\y=0$\ .			\cr}
		\eqn\parchoice$$

Using the map \homtwotry\ with \parchoice\ at the infinity of $\O^3$,
a space $\O P^3$ may be defined as $\O^3\cup\O P^2$. One may show that also
higher-dimensional octonionic spaces may be constructed once the
requirement that the maps $S^{8n-1}\!\rightarrow\!\O P^{n-1}$ be
continuous fibrations is relaxed.

There is another solution to the problem of finding homogeneous
coordinates (for $\O P^2$ only) that lies in the fact
that in contrast to the case of $\O P^1$ (two transforming fields)
we now have three inequivalent linear diagrams (figure 2)
for the $S^7$ transformations. We define the associated
transformations and hence the associated partial equivalence classes
patchwise, with one patch for each diagram,
and the complete class arises only after identifying
points in different patches via transition functions.
The set of homogeneous coordinates for $\O P^2$ is defined as follows:
$$\eqalign{&(\x_1,\y_1,\z_1)\approx(\x_1\w,\y_1\pro X\w,\z_1\pro X\w)\komma
	\quad	X={\x_1\over|\x_1|}\komma\cr
	&(\x_2,\y_2,\z_2)\approx(\x_2\pro Y\w',\y_2\w',\z_2\pro Y\w')
			\komma\quad
		Y={\y_2\over|\y_2|}\komma\cr
	&(\x_3,\y_3,\z_3)\approx(\x_3\pro Z\w'',\y_3\pro Z\w'',\z_3\w'')
			\komma\quad
		Z={\z_3\over|\z_3|}\komma\cr}\eqn\twohom$$
valid for $\x\neq 0$, $\y\neq 0$ and $\z\neq 0$ respectively (the statements
$\x_n\is 0$ \etc\ are independent of the subscripts).
The overlap relations between the three patches are the transformations
needed to go from one diagram to another, \eg\
$$
\eqalign{&\x_2=\x_1\komma\cr
	&\y_2=\y_1\komma\cr
	&\z_2=((\z_1X_1^*)(X_1Y_1^*))Y_1\komma}
\eqn\homoverlap
$$
applying for $\x\!\neq\!0$, $\y\!\neq\!0$ (\ie\ in the overlap of the
regions where the charts $1$ and $2$ are valid), creating a link between
$\z$ and $\y$ instead of the one between
$\z$ and $\x$. In this way, any of the three diagrams is related to
any other in the region where both apply. \homoverlap\ is defined such that
partial equivalence classes are transformed into each other.
A consistency check is given
by going around the closed loop in the ``diagram of diagrams'' (figure 3).
We come to the same variables modulo an $S^7$ transformation, i.e.
we stay within a given equivalence class.
The transformations \homoverlap\ between diagrams can be modified to
include also an arbitrary $S^7$ transformation, since it does not
alter the ``link invariants'' of \twocharts. It can be shown that
no such transformations may be imposed to remove the residual
$S^7$ phase obtained from the loop in figure 3.
This means that \homoverlap\ only gives well-defined transition functions
between partial equivalence classes, and not between the homogeneous
coordinates. $(\x_i,\y_i,\z_i)$ with $|\x_i|^2+|\y_i|^2+|\z_i|^2=1$
for $1\leq i\leq 3$ do not define coordinate patches on the
twentythree-sphere.
This is consistent with the point of view that the discontinuity
of \homtwotry\ seen as a map $S^{23}\!\rightarrow\!\O P^2$ is of
topological nature and can not be smeared out.
An attempt to generalize the procedure to $\O P^3$ fails. There are 16
inequivalent diagrams transforming four fields (12 linear and 4 star-shaped).
Any closed curve in the diagram of diagrams here leads to an inconsistency.
It can be shown that this property, forcing also the diagram of diagrams to
be a tree diagram, excludes some overlaps that would have been needed
in order to define $\O P^3$ in this way.

We hope to find applications to the homogeneous coordinates
\homtwotry, \twohom.
It should be possible to realize $E_6\approx SL(3;\O)$ [\Freudenthal,
\Sudbery]
on them in a ``spinor-like'' manner, much like $SO(10)\approx SL(2;\O)$
acts on its 16-dimensional spinor representations that play the role of
homogeneous coordinates for $\O P^1$ (see the following section).
That would open for for a twistor transform
[\JordanProj,\twistor-
	\jordanmech]
for elements in $J_3(\O)$
(the exceptional Jordan algebra of 3$\times$3 hermitiean octonionic
matrices [\JordanJordan-
	\Albert]) with zero Freudenthal product
[\Freudenthal] -- a known
realization of $\O P^2$ [\JordanProj,\Freudenthal,\Porteous].
Then one would have a direct analogy to the twistor transform of the
masslessness condition in $SL(2;\O)$ [\tentwistor] that leads to
$\O P^1$ as the projective light-cone (see reference [\jordanmech]).

\section{Example: Twistors in ten dimensions.}

In ten-dimensional Minkowski space, the mass-shell constraint for
a bosonic particle is $P_\mu P^\mu\is\is0$. According to the isomorphism
$SO(1,9)\approx SL(2;\O)$ [\Sudbery], $P$ may be viewed as an element
in the Jordan algebra $J_2(\O)$ of 2$\times$2 hermitean octonionic matrices,
and the constraint becomes that of scale-invariant idempotency [\jordanmech]
$$P^2=P\tr P\fullstop\eqn\ptwozero$$
This is a well known realization of $\O P^1$ [\Freudenthal,\Schafer].
The Lorentz group is the structure group of $J_2(\O)$ [\Schafer,\Sudbery].
The two rows (or columns) in $P$ fullfilling \ptwozero\ contain
the two charts \onecharts\ (up to a real scale).
This makes it possible to perform a {\it twistor transform}
[\twistor-
	\jordanmech], which amounts to
a change of the parametrization of $\O P^1$
from the Jordan algebra one to homogeneous coordinates. In $SL(2;\O)$
language, the correspondence reads ($\lambda=[\x,\y]^t$)
$$P=\lambda\lambda^\dagger=\left[\matrix{\x\x^*&\x\y^*\cr
					\y\x^*&\y\y^*\cr}\right]
			\komma\eqn\twistortrans$$
where we immediately recognize the homogeneous coordinates and
the two charts \onecharts\ in the rescaled columns.
The similarity transformations on the
homogeneous coordinates are the $S^7$ transformations of \onehommod\
(and {\it not} the traditional transformations where the components
of $\lambda$ are subject to right multiplication with the same
parameter). The scheme may be described $SO(1,9)$-covariantly,
demanding a two-component current in a spinor representation
[\tentwistor], which provides a covariant solution replacing
\onehommod\ of the singularity in the current. We expect something
similar to be possible for the case of $J_3(\O)$ described in
the previous section. The treatment of supersymmetric particles
[\supertwistor,\htwistor,\tentwistor]
introduces fermions into $\J$ along the lines described earlier,
but that falls outside the scope of this paper.

We apologize for not referring to many important papers concerning
twistors -- we have limited ourselves to contributions strictly
relevant to the division algebra twistor program.

\section{Relation to Malcev algebras.}

A Lie algebra $\L$ with antisymmetric product $[x,y]$ fulfills the
Jacobi identities
$$J(x,y,z)\equiv[[x,y],z]+[[y,z],x]+[[z,x],y]=0\fullstop\eqn\jacobizero$$
for all elements $x,y,z\inn\L$. $J$ is by definition
alternating, \ie\
completely antisymmetric in the arguments. Note the analogy with
the associator \associator\ of an alternative algebra.
In the same way as the concept of associative algebras can be weakened
to alternativity, leading to alternative algebras, including the
octonionic algebra $\O$, the Jacobi identities may be relaxed in favour
of a weaker version,
$$J(x,y,[x,z])=[J(x,y,z),x]\fullstop\eqn\malcevid$$
These are the Malcev identities, and an algebra fulfilling them
is called a {\it Malcev algebra} [\Malcev,\KuzminII].
In view of the analogy of $J$ to the associator, the correspondence
of the Malcev identities is the Moufang identity \Mouff.
The analogy goes further: the only central simple non-Lie Malcev
algebra is the commutator algebra of imaginary octonions [\KuzminII].

Malcev algebras have been considered for physical applications
in connection with Malcev-Kac-Moody [\OsipovCentral,\OsipovSugawara]
and related $N\is 8$ superconformal algebras [\softalg,\Defever].
In this context, they have the disadvantage that they cannot be
realized in terms of Poisson brackets or commutators, since then
\jacobizero\ automatically is fulfilled.

Consider the $S^7$ algebra in the form \didj, and evaluate the
right hand side at the north pole $X\is 1$ (or at any other specific $X$).
The so obtained algebra is then a Malcev algebra.
We see that what distinguishes our $S^7$ algebra from this Malcev
algebra is the transformation of the parameter field occurring in
the $X$-product, see \Kscalar, or more generally, the transformation of
the path products. It is exactly this associator term that cancels
the Malcev $J$. Note that $J$ for the octonionic commutator algebra is [\Zorn]
$$J(x,y,z)=6[x,y,z]\fullstop\eqn\Ojacobi$$
In tensor formalism, the equation responsible for the cancellation is
$$\d_iT_{jkl}(X)=2T_{mi[j}(X)T_{kl]m}(X)\equiv 2R_{ijkl}(X)
			\eqn\torsiontrans$$
(which is equivalent to the zero-curvature condition \zerocurv),
$R$ being the completely antisymmetric $X$-associator that at the
north pole reduces to $\rho$ of \commass, so that
$$J(\d_i,\d_j,\d_k)=2[T_{ijl}(X)\d_l,\d_k]+\hbox{\it cycl.}=
	4T_{ijl}(X)T_{lkm}(X)\d_m-2(\d_kT_{ijl})(X)\d_l+\hbox{\it cycl.}=0
		\fullstop\eqn\jiszero$$
Omitting the last term gives the Malcev algebra.

\chapter{Seven-sphere Kac-Moody algebra.}

\section{Current algebra and Schwinger terms.}

A Lie algebra $\L$ may be lifted to a Kac-Moody algebra
[\Moody,\GoddardOlive] $\widehat\L$ consisting of the mappings
$S^1\!\!\rightarrow\!\!\L$ by applying the Lie product pointwise on the circle.
The interesting feature of this structure is of course that it allows
for non-trivial central extensions, or Schwinger terms [\Schwinger].
The classical version of our ``$S^7$ Kac-Moody algebra'', $\widehat{S^7}$,
is therefore trivial --- in a conformal field theory language we
simply have
$$
\wickcontract{\J_\a(z)}{\J_\b(\z)}={1\over z-\z}\J_{[\a,\b]_\X}
	\komma\eqn\classKM
$$
ignoring potential normal ordering terms.

Now we have a set of structure functions (the torsion tensor) that
varies over $S^7$, so it can be expected that the Schwinger terms
can exhibit a similar behaviour (this is why we in the generic case avoid
the notion of ``central extensions''). This is easily demonstrated.

Take the simplest realization, where only the parameter field $\x$ and
its momentum $\p$ transform, and the current is
$$\J=\{\p^*\x\}\fullstop	\eqn\KMone$$
We use a conformal field theory language, \ie\ let the fields be
holomorphic in a complex variable $z$, and postulate the fundamental
correlator
$$
\wickcontract{\x_o(z)}{\p_{o'}(\z)}=
{\hbar[o^*o']\over z-\z}\fullstop\eqn\fundcorr
$$
Then the correlator of two currents is easily evaluated as
$$
\wickcontract{\J_\a(z)}{\J_\b(\z)}= -{8\hbar^2[\a\b]\over(z-\z)^2}+
	{\hbar\over z-\z}\J_{[\a,\b]_\X}\fullstop\eqn\KMonecorr
$$
In this simplest example, the Schwinger term is obviously a central
extension. This is not so in general. If we stay with the free fields
$\x$ ($X=\x/|\x|$) and $\p$ and the correlator \fundcorr,
 but let the current get a ``quantum correction'' according to
$$\J=\{\p^*\x\}+\sigma\hbar X^*\partial X\komma\quad
	\sigma\in\R\komma\eqn\KMonemod$$
the algebra acquires a non-central extension:
$$
\wickcontract{\J_\a(z)}{\J_\b(\z)}= -{(8+2\sigma)\hbar^2[\a\b]\over(z-\z)^2}+
	{\sigma\hbar^2\over z-\z}(X^*\partial X)_{[\a,\b]_\X}+
	{\hbar\over z-\z}\J_{[\a,\b]_\X}\fullstop\eqn\KMonemodcorr
$$
The currents of equation \KMonemod\ for different values of $\sigma$
carry the same field content and the algebras \KMonemodcorr\ can
therefore be taken as equivalent up to a (quantum) redefinition
of the current. The central extension of \KMonecorr\ can be taken
as a representative of this class.

Let us now turn to currents constructed from several octonions,
as in \totalJ, and for simplicity we first treat the case
\lmgen\ of two transforming fields.
It is easily seen that any double contraction (giving rise to
Schwinger terms) must take place between identical terms in the
two currents, since these are linear in momenta. The Schwinger terms
are therefore ``additive'' -- introducing new terms in $\J$ corresponding
to new transforming fields only gives rise to extra Schwinger terms
arising from double contractions of these terms with themselves.
The second term of \lmgen\ gives a field dependent double
contraction, so that the
quantum current algebra now takes the form (from now on, $\hbar$ is
suppressed)
$$
\wickcontract{\J_\a(z)}{\J_\b(\z)}= -{16[\a\b]\over(z-\z)^2}+
	{4\over z-\z}(X^*\partial X)_{[\a,\b]_\X}+
	\J_{[\a,\b]_\X}\fullstop\eqn\qtwocorr
$$
Here we have a field dependent Schwinger term.
Notice that the anomaly is in the one-parameter class of \KMonemodcorr\
with $\sigma\is 4$, so that adding a quantum correction
$-4X^*\partial X$ to $\J$ gives back \KMonecorr.
In fact, the second transforming octonion has not changed
the numerical coefficient of the central extension.
This result generalizes to any number of fields -- adding a quantum
correction to \totalJ\ to obtain
$$\J_\a=-\sum_k[\p_k^*(\r_k\Pro{\P[\r_k]}\a)]+
	4\sum_{k\neq 0}[X_{k-1}^*\partial X_{k-1}\a_k]\komma\eqn\qtotalJ$$
with $\a_k$ defined as in \alttotalJ,
gives back the correlator \KMonecorr. We obtain the surprising result,
contrasting to the situation in (Lie) Kac-Moody algebras,
that the coefficient of the central extension is {\it unique} up to
redefinitions of the current. This statement of course holds
given that the field content is that of section 2.6. A more
general uniqueness proof will have to wait for the general
representation theory to be completed.

Any conjugate pair of fermions (at an endpoint of the tree diagram)
contributes to the Schwinger term with the same term as a boson in the
same position would have done, but with opposite sign, just as for
Lie Kac-Moody algebras. If the fermion is self-conjugate, the coefficient
is divided by two. The forms of the quantum corrections needed
to render the algebra extension central are completely analogous.

\section{BRST operator. ``Anomaly cancellation''.}

In this section we would like to address the question of anomaly
cancellation: under what circumstances is the Schwinger term
``quantum mechanically consistent'', \ie\ when is the BRST operator
quantum mechanically nilpotent, and what actual exact form of the
Schwinger term is needed? This question will be of relevance if
the algebra is considered as an algebra of gauge constraints,
\eg\ in some twistor string model. It will be shown that quantum mechanical
consistency is compatible with one member of the class of anomalies
obtained above.

The first thing to do is to examine how to construct a (classical)
BRST operator for the $S^7$ algebra with field-dependent structure
functions. This turns out to be extremely simple. The BRST operator
takes the same form as for a Lie algebra, namely
$$\Q=c^i\J_i-{T_{ij}}^k(X)c^ic^jb_k\komma\eqn\BRSTop$$
where $b_i$ and $c^i$ are fermionic ghosts with $b_i(z)c^j(\z)\sim
\delta_i^j(z-\z)^{-1}$. Higher order ghost terms are not present
since the Jacobi identities hold, due to \torsiontrans. This makes
BRST analysis quite manageable.

Then, turning to $\widehat{S^7}$ and the quantum algebra,
a somewhat lengthy calculation shows
that the current must obey \KMonemodcorr\ with $\sigma\is 8$ in order
for $\Q$ (with \BRSTop\ as the BRST charge density)
to be quantum mechanically nilpotent:
$$\Q^2=0\quad\Longleftrightarrow
	\wickcontract{\J_\a(z)}{\J_\b(\z)}\ = \
	-{24[\a\b]\over(z-\z)^2}+
	{1\over z-\z}(\J + 8X^*\partial X)_{[\a,\b]_\X}
		\fullstop\eqn\BRSTcond
$$
We have thus demonstrated the non-trivial fact that $\Q$ {\it may} be
nilpotent, and that $\widehat{S^7}$ may be used as a gauge algebra.
Normally, one would have expected $\Q^2\is 0$ to put a constraint on the
number of transforming octonionic fields, but that is not the case
at hand. Instead one is permitted, for any field content, to adjust the
numerical coefficient of $X^*\partial X$ in $\J$ in order to fulfil
that relation\foot{One can however imagine that other covariance
properties, for example Lorentz symmetry, puts a restriction on the
quantum corrections, so that \BRSTcond\ becomes predictive for the
field content.}.

One may remark that if one restricts oneself to the plain generators
of \totalJ\ without quantum corrections, the BRST charge is nilpotent
for the (linear) diagram of three bosonic fields, with the parameter
field in the middle. We suspect that it is more than a coincidence
that this is the number of fields transforming under $\widehat{S^7}$
in the string twistor model of [\Berkovitsstring], where one has the
pair of octonions making the twistor variable $\lambda$ (see section 2.8)
and the ghosts for the eight supercharges of the associated superconformal
algebra [\Berkovitsstring,\Eightconf,\hconf]. We have not carried out the
detailed analysis, however.

In an ordinary algebra with structure {\it constants}, one can let $\Q$
act on the $b$ field to obtain a modified, BRST-exact, current
$\tilde\J$ containing the ghost fields. If $\Q^2\is 0$, then the algebra
of $\tilde\J$ is non-anomalous. This is not so here. Since the Jacobi
identities hold, it is easy to show that a Poisson bracket of two
BRST-exact operators is BRST-exact, but for the $S^7$ algebra
one obtains
$$\{\{b_i,\Q\},\{b_j,\Q\}\}=\{2{T_{ij}}^k(X)b_k,\Q\}\neq
	2{T_{ij}}^k(X)\{b_k,\Q\}\fullstop\eqn\noghostJ$$
This specific subset of the BRST-exact operators does not close to an
algebra. It seems that one has to conclude that the $S^7$ or
$\widehat{S^7}$ ghosts do not come in an $S^7$ 
representation. This is also confirmed by an attempt to construct a
representation (other than scalar) for imaginary octonions, which turns
out to be impossible. This was already hinted at when dealing with
tensor composition of spinors in section 2.3.

\section{Sugawara construction.}

The similarity of $S^7$ to a group manifold and the results of
Osipov [\OsipovSugawara] for Malcev-Kac-Moody algebras
let us expect that a Sugawara construction
is possible also for $\widehat{S^7}$. This is not difficult to
verify for the simplest current $\J=\{\p^*\x\}$ and the associated
energy-momentum tensor
$$
\eqalign{
L 	&= - {1\over 8} : \J_j \J_j : \cr
	&= {7\over 8} [ \x^* \partial \p - \partial\x^* \p]
		+ {1\over 8} [\x^*\p e_j ] [\x^*\p e_j ]\ \ , \cr
}
\eqn\sugx
$$
where we have normal-ordered the currents in the standard way:
$$
 : \J_j(\z) \J_k(\z) :\ = \lim_{z\rightarrow \z}
	( \J_j(z) \J_k(\z) - {56\over (z-\z)^2} )\ .
\eqn\normalone
$$
In order to generalize this result to currents with arbitrary
field content, one has to be careful about normal ordering.
Up to this point we have implicitly assumed free-field
normal ordering on the right-hand side of the current-current
commutation relations. Even though the torsion tensor
$T_{ijk}(X)$ commutes with $\J_k$, the product of those
two operators obeys
$$
\eqalign{
T_{ijk}(X)\J_k &=\ : T_{ijk}(X)\ \J_k :
-\ 8T_{ijk}(X)[X^* \partial X e_k]\cr
&=\ : T_{ijk}(X)\J_k : +\ T_{ikl}(X) \partial_z T_{jkl}(X) \  ,\cr
}
\eqn\normaltwo
$$
where the left-hand side free-field normal ordered and
the first term on the right-hand side is current normal ordered.
It is useful to employ the technology described by
Bais, Bouwknegt, Surridge and Schoutens [\Holland] to show the
following result:
for currents that obey
$$
\wickcontract{\J_i(z)\ }{ \J_j(\z)} = {k\delta^{ij} \over (z-\z)^2} +
{2\over z-\z}  : T_{ijk}(X) \bigl( \J_k - 8  [X^* \partial X e_k]\bigr )(\z):
\ ,
\eqn\currope
$$
the Sugawara energy-momentum tensor is given by
$$
L ={1\over 2k-24} : \J_j \J_j :
\eqn\sugj
$$
and has a central charge of
$$
c = {7k \over k-12} \ ,
\eqn\sugjcent
$$
in accordance with the known results for Kac-Moody algebras.
\sugj\ and \sugjcent\ are identical to Osipov's
formulas [\OsipovSugawara]. We note that in contrast to
the Kac-Moody case, the requirement that
$\J_i(z) + \alpha [X^* \partial X e_i](z)$ should transform
like a dimension 1 current under the action of the Sugawara
energy-momentum tensor fixes only the constant $\alpha$.
There is a three-parameter family of candidate Sugawara
tensors which satisfy this condition. The operator product
of $L(z)$ with itself determines the parameters. There are
two solutions: \sugj\ for any $k$ and another solution
which would requires a complex value of $k$. The Sugawara
construction is therefore unique.

\chapter{Summary and discussion.}

We have discussed several aspects of the seven-sphere algebra and some
related topics. We find it somewhat surprising that this algebra has
recieved so little attention in the mathematical literature (compared to
Malcev algebras), in spite of the fact that the parallelizability property
has been known for a long time, and the simplicity of the argument
in section 2.1.

{}From a physical point of view, the $S^7$ algebra provides a natural
generalization of the Lie algebra concept. We have demonstrated how
it can be handled when arising as a gauge algebra of constraints (BRST
procedure) and how it can be used as a generalized Kac-Moody-Lie algebra.
For this last case, some unexpected features of the Schwinger term occur,
distinguishing it from ordinary Kac-Moody-Lie algebras. The feasability
of a BRST procedure involving field-dependent structure functions and
anomalies is not {\it a priori} ascertained, but has been demonstrated.
The class of physical models closest in our minds for this kind of
symmetry is string twistor theories. Different versions have been
formulated, but at least one of them possesses an $S^7$ Kac-Moody
gauge symmetry [\Berkovitsstring]. Superstring twistors involve
a super-extension to an $N\is8$ superconformal algebra
[\Berkovitsstring-
\hconf],
and we hope that it will be possible to give a similar treatment of
the super-extensions to the one presented
for $S^7$ in the present paper. Especially the problem of anomaly
cancellation may gain some insight from our results.

A part of the structure of $S^7$ we have treated only fragmentarily is
representation theory. We would like to return to that question later.
It is not immediately clear even how to define a representation.
We have quite strong feelings, though, that the spinorial representations
and the adjoint, as described in this paper, in some sense are the
only ones allowed, and that the spinor representation is the only one
to which a variable freely can be assigned.

Most of this paper has been written without specific aim at physical
applications, mostly because we felt that our mathematical understanding
of the algebras we were dealing with in ten-dimensional superstring models
was dragging behind. This means that some sections may be of little
interest when studying a specific physical problem. On the other hand,
we find some of our byproducts, \eg\ those concerning infinitesimal generation
of the octonionic Hopf map, quite appealing by themselves.

\vskip12pt
\ack We would like to thank Lars Brink, Nathan Berkovits,
Volodia Fock, Viktor Ogievetsky and Peter West for valuable discussions.

\refout
\endpage

\vbox to 20cm{
	\hbox{\centerline{\twelverm Figure Caption}}
	\vfil
	\hbox{\hskip4.5cm\epsffile{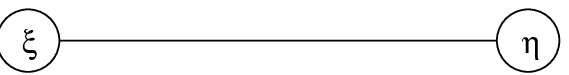}\hfil}
	\vfil
	\hbox{\centerline{Figure 1. Diagram for $\O P^1$.}}
	\vfil
	\epsffile{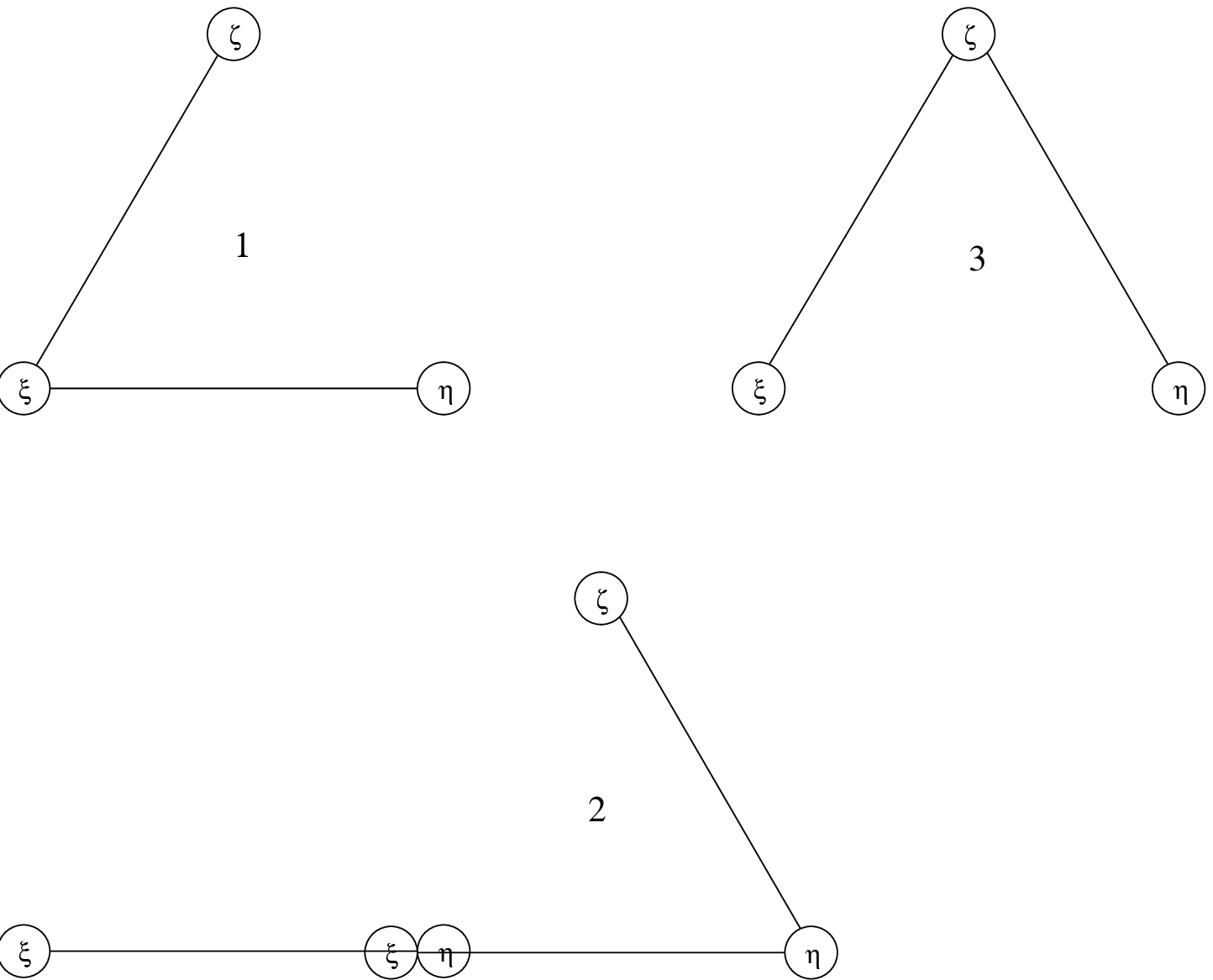}
	\vfil
	\hbox{\centerline{Figure 2. Diagrams for $\O P^2$.}}
		}
\endpage
\vbox to 10cm{
	\vfil
	\hbox{\hskip3cm\epsffile{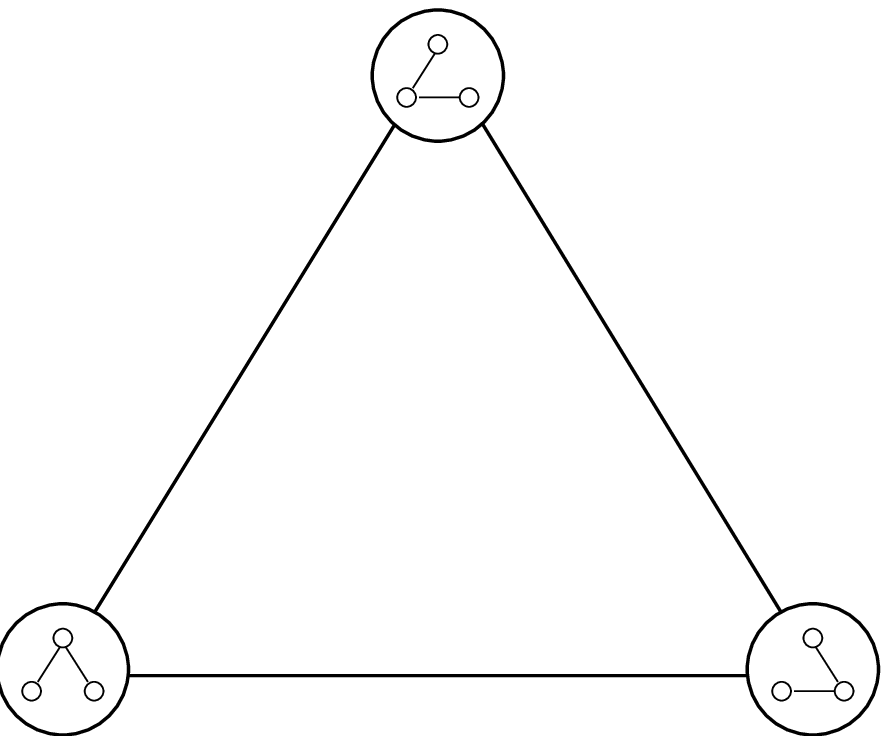}\hfil}
	\vfil
	\hbox{\centerline{Figure 3. Diagram of diagrams for $\O P^2$.}}
		}
\end